# Mechanizing Coinduction and Corecursion
# in
# Higher-order Logic

*Lawrence C. Paulson*
Computer Laboratory, University of Cambridge

## Abstract

A theory of recursive and corecursive definitions has been developed in higher-order logic (HOL) and mechanized using Isabelle. Least fixedpoints express inductive data types such as strict lists; greatest fixedpoints express coinductive data types, such as lazy lists. Well-founded recursion expresses recursive functions over inductive data types; corecursion expresses functions that yield elements of coinductive data types. The theory rests on a traditional formalization of infinite trees.

The theory is intended for use in specification and verification. It supports reasoning about a wide range of computable functions, but it does not formalize their operational semantics and can express noncomputable functions also. The theory is illustrated using finite and infinite lists. Corecursion expresses functions over infinite lists; coinduction reasons about such functions.

**Key words.** Isabelle, higher-order logic, coinduction, corecursion

# Contents

# 1 Introduction

Recursive data structures, and recursive functions over them, are of central interest in Computer Science. The underlying theory is that of inductive definitions [2]. Much recent work has focused on formalizing induction principles in type theories. The type theory of Coq takes inductive definitions as primitive [8]. The second-order $\lambda$-calculus (known variously as System $F$ and $\lambda 2$) can express certain inductive definitions as second-order abstractions [13].

Of growing importance is the dual notion: *coinductive* definitions. Infinite data structures, such as streams, can be expressed coinductively. The dual of recursion, called *corecursion*, can express functions involving coinductive types.

Coinduction is well established for reasoning in concurrency theory [20]. Abramsky's Lazy Lambda Calculus [1] has made coinduction equally important in the theory of functional programming. Milner and Tofte motivate coinduction through a simple proof about types in a functional language [21]. Tofte has proved the soundness of a type discipline for polymorphic references by coinduction [32]. Pitts has derived a coinduction rule for proving facts of the form $x \sqsubseteq y$ in domain theory [28].

There are many ways of formalizing coinduction and corecursion. Mendler [19] has proposed extending $\lambda 2$ with inductive and coinductive types, equipped with recursion and corecursion operators. More recently, Geuvers [12] has shown that coinductive types can be constructed from inductive types, and vice versa. Leclerc and Paulin-Mohring [17] investigate various formalizations of streams in the Coq system. Rutten and Turi survey three other approaches [29].

Church's higher-order logic (HOL) is perfectly adequate for formalizing both inductive and coinductive definitions. The constructions are not especially difficult. A key tool is the Knaster-Tarski theorem, which yields least and greatest fixedpoints of monotone functions. Trees, finite and infinite, are represented as sets of nodes. Recursion is derivable in its most general form, for arbitrary well-founded relations. Corecursion and coinduction have straightforward definitions.

Compared with other type theories, HOL has the advantage of being simple, stable and well-understood. But $\lambda 2$ and Coq have an inbuilt operational semantics based on reduction, while a HOL theory can only suggest an operational interpretation. HOL admits non-computable functions, which is sometimes advantageous and sometimes not.

Higher-order logic is extremely successful in verification, mainly hardware verification [7]. The HOL system [14] is particularly popular. Melham has formalized and mechanized a theory of inductive definitions for the HOL system [18]; my work uses different principles to lay the foundation for mechanizing a broader class of definitions. Some of the extensions (mutual recursion, extending the language of type constructors) are also valid in set theory [26]. One unexplored possibility is that trees may have infinite branching.

A *well-founded* (WF) relation $\prec$ admits no infinite descents $\cdots \prec x_n \prec \cdots x_1 \prec x_0$. A data structure is WF provided its substructure relation is well-founded. My work justifies non-WF data structures, which involve infinitely deep nesting.

The paper is chiefly concerned with coinduction and corecursion. §2 briefly introduces Isabelle and its formalization of HOL. §3 describes the least and greatest fixedpoint operators. §4 presents the representation of infinite trees. §5 considers finite lists and other WF data structures. §6 concerns lazy lists, deriving coinduction principles for type-checking and equality. §7 introduces and derives corecursion, while §8 presents examples. Finally, §9 gives conclusions and discusses related work.

# 2 HOL in Isabelle

The theory described below has been mechanized using Isabelle, a generic theorem prover [25]. Isabelle implements several object-logics: first-order logic, Zermelo-Frænkel set theory, Constructive Type Theory, higher-order logic (HOL), etc.

Isabelle has logic programming features, such as unification and proof search. Every Isabelle object-logic can take advantage of these. Isabelle's rewriter and classical reasoning package can be used with logics having the appropriate properties.

For this paper, we may regard Isabelle simply as an implementation of higher-order logic. There are many others, such as TPS [4], IMPS [10] and the HOL system [14]. I shall describe the theory of recursion in formal detail, to facilitate its mechanization in any suitable system.

## 2.1 Higher-order logic as an meta-logic

Isabelle exploits the power of higher-order logic on two levels. At the meta-level, Isabelle uses a fragment of intuitionistic HOL to mechanize inference in various object-logics. One of these object-logics is classical HOL.

The meta-logic, as a fragment of HOL, is based upon the typed $\lambda$-calculus. It uses $\lambda$-abstraction to formalize the object-logic's binding operators, such as $\forall x\ \phi$, $\prod_{x\in A} B$, $\epsilon x.\phi$ and $\bigcup_{x\in A} B$, in the same manner as Church did for higher-order logic [6]. The approach is fully general; each binding operator may involve any fixed pattern of arguments and bound variables, and may denote a formula, term, set, type, etc. In a recent paper [24], I discuss variable binding with examples.

Quantification in the meta-logic expresses axiom and theorem schemes. Binding operators typically involve higher-order definitions. The normalization theorem for natural deduction proofs in HOL can be used to justify the soundness of Isabelle's representation of the object-logic. I have done a detailed proof for the case of intuitionistic first-order logic [23]; the argument applies, with obvious modifications, to any formalization of a similar syntactic form.

## 2.2 Higher-order logic as an object-logic

Isabelle/HOL, Isabelle's formalization of higher-order logic, follows the HOL system [14] and thus Church [6]. The connectives and quantifiers are defined in terms of three primitives: implication ($\rightarrow$), equality ($=$) and descriptions ($\epsilon x.\phi$). Isabelle emphasizes a natural deduction style; its HOL theory derives natural deduction rules for the connectives and quantifiers from their obscure definitions.

The *types* $\sigma$, $\tau$, ..., are simple types. The type of truth values is called `bool`. The type of individuals plays no role in the sequel; instead, we use types of natural numbers, products, etc. Church used subscripting to indicate type, as in $x_\tau$; Isabelle's notation is $x :: \tau$. Church wrote $\tau\sigma$ for the type of functions from $\sigma$ to $\tau$; Isabelle's notation is $\sigma \Rightarrow \tau$. Nested function types may be abbreviated:

$$(\sigma_1 \Rightarrow \cdots (\sigma_n \Rightarrow \tau) \cdots) \quad \text{as} \quad [\sigma_1, \ldots, \sigma_n] \Rightarrow \tau$$

ML-style polymorphism replaces Church's type-indexed families of constants. *Type schemes* containing type variables $\alpha$, $\beta$, ..., represent families of types. For example, the quantifiers have type $(\alpha \Rightarrow \texttt{bool}) \Rightarrow \texttt{bool}$; the type variable $\alpha$ may be instantiated to any suitable type for the quantification.[1] As in ML, type operators have a postfix notation. For example, $(\sigma)\,\texttt{set}$ is the type of sets over $\sigma$, and $(\sigma)\,\texttt{list}$ is the type of lists over $\sigma$. The parentheses may be omitted if there is no ambiguity; thus `nat set set` is the type of sets of sets of natural numbers.

The *terms* are those of the typed $\lambda$-calculus. The *formulae* are terms of type `bool`. They are constructed from the logical constants $\wedge$, $\vee$, $\rightarrow$ and $\neg$. There is no $\leftrightarrow$ connective; the equality $\phi = \psi$ truth values serves as a biconditional. The *quantifiers* are $\forall$, $\exists$ and $\exists!$ (unique existence).

## 2.3 Further Isabelle/HOL types

Isabelle's higher-order logic is augmented with Cartesian products, disjoint sums, the natural numbers and arithmetic. Isabelle theories define the appropriate types and constants, and prove a large collection of theorems and derived rules.

The *singleton* type `unit` possesses the single value $() :: \texttt{unit}$.

The *Cartesian product* type $\sigma \times \tau$ possesses as values ordered pairs $(x, y)$, for $x$ of type $\sigma$ and $y$ of type $\tau$. We have the usual two projections, as well as the eliminator `split`:

$$\begin{aligned} \texttt{fst} &:: (\alpha \times \beta) \Rightarrow \alpha \\ \texttt{snd} &:: (\alpha \times \beta) \Rightarrow \beta \\ \texttt{split} &:: [[\alpha, \beta] \Rightarrow \gamma,\ \alpha \times \beta] \Rightarrow \gamma \end{aligned}$$

Operations over pairs can be couched in terms of `split`, which satisfies the equation $\texttt{split}\ c\ (a, b) = c\ a\ b$.

The *disjoint sum* type $\sigma + \tau$ possesses values of the form $\texttt{Inl}\ x$ for $x :: \sigma$ and $\texttt{Inr}\ y$ for $y :: \tau$. The eliminator, `sum_case`, is similar in spirit to `split`:

$$\texttt{sum_case} :: [\alpha \Rightarrow \gamma,\ \beta \Rightarrow \gamma,\ \alpha + \beta] \Rightarrow \gamma$$

The eliminator performs case analysis on its first argument:

$$\begin{aligned} \texttt{sum_case}\ c\ d\ (\texttt{Inl}\ a) &= c\ a \\ \texttt{sum_case}\ c\ d\ (\texttt{Inr}\ b) &= d\ b \end{aligned}$$

The operators `split` and `sum_case` are easily combined to express pattern matching over more complex arguments. The compound operator

$$\texttt{sum_case}\,(\texttt{sum_case}\ c\ d)\,(\texttt{split}\ e)\ z \tag{1}$$

[1] In Isabelle, type variables are classified by sorts in order to control polymorphism, but this need not concern us here.

analyses an argument $z$ of type $(\alpha+\beta)+(\gamma\times\delta)$. This is helpful to implementors writing packages to support such data types. On the other hand, expressions involving `split` and `sum_case` are hard to read. Therefore Isabelle allows limited pattern matching and case analysis in definitions. A `case` syntax can be used for `sum_case` and similar operators. In a binding position, a pair of variables stands for a call to `split`. These constructs nest. We can express the operator (1) less concisely but more readably by

$$\begin{array}{ll} \texttt{case}\ z\ \texttt{of}\ \ \texttt{Inl}\ z' & \Rightarrow (\texttt{case}\ z'\ \texttt{of}\ \ \texttt{Inl}\ x \Rightarrow c\ x \\ & \qquad\qquad\quad \mid \texttt{Inr}\ y \Rightarrow d\ y) \\ \quad\mid \texttt{Inr}(v,w) \Rightarrow e\ v\ w & \end{array}$$

The *natural number* type, `nat`, has the usual arithmetic operations $+$, $-$, $\times$, etc., of type $[\texttt{nat},\texttt{nat}] \Rightarrow \texttt{nat}$. The successor function is $\texttt{Suc} :: \texttt{nat} \Rightarrow \texttt{nat}$. Definitions involving the natural numbers can use the `case` construct, with separate cases for zero and successor numbers. Isabelle also supports a special syntax for definition by primitive recursion. This paper presents definitions using the most readable syntax available in Isabelle, occasionally going beyond this.

## 2.4 Sets in Isabelle/HOL

Set theory in higher-order logic dates back to *Principia Mathematica*'s theory of classes [33]. Although sets are essentially predicates, Isabelle/HOL defines the type $\alpha\,\texttt{set}$ for sets over type $\alpha$. Type $\alpha\,\texttt{set}$ possesses values of the form $\{x \mid \phi x\}$ for $\phi :: \alpha \Rightarrow \texttt{bool}$. The eliminator is membership, $\in :: [\alpha, \alpha\,\texttt{set}] \Rightarrow \texttt{bool}$. It satisfies the equations

$$\begin{aligned} (a \in \{x \mid \phi x\}) &= \phi a \\ \{x \mid x \in A\} &= A \end{aligned}$$

Types distinguish this set theory from axiomatic set theories such as Zermelo-Frænkel. All elements of a set must have the same type, and formulae such as $x \notin x$ and $x \in y \wedge y \in z \to x \in z$ are ill-typed. For each type $\alpha$, there is a universal set, namely $\{x \mid \texttt{True}\}$.

Many set-theoretic operations have obvious definitions:

$$\begin{aligned} \forall_{x\in A}\,\psi &\equiv \forall x\,(x \in A \to \psi) \\ \exists_{x\in A}\,\psi &\equiv \exists x\,(x \in A \wedge \psi) \\ A \subseteq B &\equiv \forall_{x\in A}\, x \in B \\ A \cup B &\equiv \{x \mid x \in A \vee x \in B\} \\ A \cap B &\equiv \{x \mid x \in A \wedge x \in B\} \end{aligned}$$

We have several forms of large union: the bounded union $\bigcup_{x\in A} B$, the unbounded union $\bigcup_x B$, and the union of a set of sets, $\bigcup S$. Their definitions, and those of the corresponding intersection operators, are straightforward:

$$\bigcup_{x\in A} B \equiv \{y \mid \exists_{x\in A}\, y \in B\}$$
$$\bigcap_{x\in A} B \equiv \{y \mid \forall_{x\in A}\, y \in B\}$$
$$\bigcup_{x} B \equiv \bigcup_{x\in\{x\mid \mathtt{True}\}} B$$
$$\bigcap_{x} B \equiv \bigcap_{x\in\{x\mid \mathtt{True}\}} B$$
$$\bigcup S \equiv \bigcup_{x\in S} x$$
$$\bigcap S \equiv \bigcap_{x\in S} x$$

Our definitions will frequently refer to the range of a function, and to the image of a set over a function:

$$\mathtt{range}\, f \equiv \{y \mid \exists x\, y = f\ x\}$$
$$f \text{ “ } A \equiv \{y \mid \exists_{x\in A}\, y = f\ x\}$$

For reasoning about the set operations, I prefer to derive natural deduction rules. For example, there are two introduction rules for $A \cup B$:

$$\frac{x \in A}{x \in A \cup B} \qquad \frac{x \in B}{x \in A \cup B}$$

The corresponding elimination rule resembles disjunction elimination.

The Isabelle theory includes the familiar properties of the set operations. Isabelle's classical reasoner can prove many of these automatically. Examples:

$$\bigcap(A \cup B) = \bigcap A \cap \bigcap B$$
$$\left(\bigcup_{x\in C} A\ x \cup B\ x\right) = \bigcup(A \text{ “ } C) \cup \bigcup(B \text{ “ } C)$$

## 2.5 Type definitions

In Gordon's HOL system, a new type $\tau$ can be defined from an existing type $\sigma$ and a predicate $\phi :: \sigma \Rightarrow \mathtt{bool}$. Each element of type $\tau$ is represented by some element $x :: \sigma$ such that $\phi x$ holds. The type definition is valid only if $\exists x\, \phi x$ is a theorem, since HOL does not admit empty types. Unicity of types demands a distinction between elements of $\sigma$ and elements of $\tau$, which can be achieved by introducing an *abstraction* function $\mathtt{abs} :: \sigma \Rightarrow \tau$. Each element of $\tau$ has the form $\mathtt{abs}\ x$ for some $x :: \sigma$ such that $\phi x$ holds. The function abs has a right inverse, the *representation* function $\mathtt{rep} :: \tau \Rightarrow \sigma$, satisfying

$$\phi(\mathtt{rep}\ y), \qquad \mathtt{abs}(\mathtt{rep}\ y) = y, \qquad \text{and} \qquad \phi x \to \mathtt{rep}(\mathtt{abs}\ x) = x.$$

As a trivial example, the singleton type unit serves as a nullary Cartesian product. It may be defined from bool by the predicate $\lambda x.\, x = \mathtt{True}$. Calling the abstraction function $\mathtt{abs_unit} :: \mathtt{bool} \Rightarrow \mathtt{unit}$, we obtain $() :: \mathtt{unit}$ by means of the definition

$$() \equiv \mathtt{abs_unit}(\mathtt{True})$$

Isabelle and the HOL system have polymorphic type systems; types may contain type variables $\alpha$, $\beta$, ... , which range over types. Type variables may also occur in terms:

$$\lambda f :: [\alpha, \beta] \Rightarrow \texttt{bool}. \exists a\, b\, (f = (\lambda x\, y.\, x = a \wedge y = b))$$

This is a predicate over the type scheme $[\alpha, \beta] \Rightarrow \texttt{bool}$. It allows us to define $\alpha \times \beta$, the Cartesian product of two types. We may then write $\sigma \times \tau$ for any two types $\sigma$ and $\tau$.

Like the HOL system, Isabelle/HOL supports type definitions. A `subtype` declaration specifies the new type name and a set expression. This set determines the representing type and the predicate over it, called $\sigma$ and $\phi$ above. Isabelle automatically declares the new type; its abstraction and representation functions receive names of the form `Abs_`$X$ and `Rep_`$X$.

# 3 Least and greatest fixedpoints

The Knaster-Tarski Theorem asserts that each monotone function over a complete lattice possesses a fixedpoint.[2] Tarski later proved that the fixedpoints themselves form a complete lattice [31]; we shall be concerned only with the least and greatest fixedpoints. Least fixedpoints yield inductive definitions while greatest fixedpoints yield coinductive definitions.

Our theory of inductive definitions requires only one kind of lattice: the collection, ordered by the relation $\subseteq$, of the subsets of a set. Monotonicity is defined by

$$\texttt{mono}\ f \equiv \forall A\, B (A \subseteq B \rightarrow f\ A \subseteq f\ B).$$

Monotonicity is generally easy to prove. The following results each have one-line proofs in Isabelle:

$$\frac{A \subseteq C \quad B \subseteq D}{A \cup B \subseteq C \cup D} \qquad \frac{A \subseteq B}{f\, ``\, A \subseteq f\, ``\, B} \qquad \frac{A \subseteq C \quad \begin{matrix}[x \in A]_x \\ \vdots \\ B\, x \subseteq D\, x\end{matrix}}{(\bigcup_{x \in A} B\, x) \subseteq (\bigcup_{x \in C} D\, x)}$$

Armed with facts such as these, it is trivial to prove that a function composed from such operators is itself monotonic.

The fixedpoint operators are called `lfp` and `gfp`. They both have type $[\alpha\,\texttt{set} \Rightarrow \alpha\,\texttt{set}] \Rightarrow \alpha\,\texttt{set}$; they take a monotone function and yield a set.

## 3.1 The least fixedpoint

The least fixedpoint operator is defined by $\texttt{lfp}\ f \equiv \bigcap\{X \mid f\ X \subseteq X\}$. Roughly speaking, `lfp` $f$ contains only those objects that must be included: they are common to all fixedpoints of $f$. The Isabelle theory proves that `lfp` is indeed the least fixedpoint of $f$:

$$\frac{f\ A \subseteq A}{\texttt{lfp}\ f \subseteq A} \qquad \frac{\texttt{mono}\ f}{\texttt{lfp}\ f = f(\texttt{lfp}\ f)}$$

[2]See Davey and Priestley for a modern discussion [9].

The fixedpoint property justifies both introduction and elimination rules for $\texttt{lfp}\ f$, assuming we already know how to construct and take apart sets of the form $f\ A$. Because $\texttt{lfp}\ f$ is the *least* fixedpoint, it satisfies a better elimination rule, namely induction. The Isabelle theory derives a strong form of induction, which can easily be instantiated to yield structural induction rules:

$$\frac{a \in \texttt{lfp}\ f \quad \texttt{mono}\ f \quad \begin{matrix}[x \in f(\texttt{lfp}\ f \cap \{x \mid \psi x\})]_x \\ \vdots \\ \psi x\end{matrix}}{\psi a}$$

The set $\texttt{List}\ A$ of finite lists over $A$ is a typical example of a least fixedpoint. Lists have two introduction rules:

$$\texttt{NIL} \in \texttt{List}\ A \qquad \frac{M \in A \quad N \in \texttt{List}\ A}{\texttt{CONS}\ M\ N \in \texttt{List}\ A}$$

The elimination rule is structural induction:

$$\frac{M \in \texttt{List}\ A \quad \psi\texttt{NIL} \quad \begin{matrix}[x \in A \quad y \in \texttt{List}\ A \quad \psi y]_{x,y} \\ \vdots \\ \psi(\texttt{CONS}\ x\ y)\end{matrix}}{\psi M}$$

The related principle of structural recursion expresses recursive functions on finite lists. See §5.1 for details of the definition of lists. Elsewhere [26] I discuss the lfp induction rule and other aspects of the lfp theory.

## 3.2 The greatest fixedpoint

The greatest fixedpoint operator is defined by $\texttt{gfp}\ f \equiv \bigcup\{X \mid X \subseteq f\ X\}$. The dual of the least fixedpoint, it excludes only those elements that must be excluded. The Isabelle theory proves that gfp is the greatest fixedpoint of $f$:

$$\frac{A \subseteq f\ A}{A \subseteq \texttt{gfp}\ f} \qquad \frac{\texttt{mono}\ f}{\texttt{gfp}\ f = f(\texttt{gfp}\ f)}$$

As with $\texttt{lfp}\ f$, the fixedpoint property justifies both introduction and elimination rules for $\texttt{gfp}\ f$. But the elimination rule is not induction; instead, a further introduction rule is coinduction.

Typically $\texttt{gfp}\ f$ contains infinite objects. The usual introduction rules, like those for NIL and CONS above, can only justify finite objects — each rule application justifies only one stage of the construction. But a single application of coinduction can prove the existence of an infinite object. Conversely, we should not expect to have a structural induction rule when there are infinite objects.

To show $a \in \texttt{gfp}\ f$ by coinduction, exhibit some set $X$ such that $a \in X$ and $X \subseteq f\ X$:

$$\frac{a \in X \quad X \subseteq f\ X}{a \in \texttt{gfp}\ f}$$

This rule is *weak* coinduction. Its soundness is obvious by the definition of gfp and does not even require $f$ to be monotonic. The set $X$ is typically a singleton or the range of some function.

For monotonic $f$, coinduction can be strengthened in various ways. The following version is called *strong* coinduction below:

$$\frac{a \in X \quad X \subseteq f\ X \cup \mathtt{gfp}\ f \quad \mathtt{mono}\ f}{a \in \mathtt{gfp}\ f}$$

An even stronger version, not required below, is

$$\frac{a \in X \quad X \subseteq f(X \cup \mathtt{gfp}\ f) \quad \mathtt{mono}\ f}{a \in \mathtt{gfp}\ f.}$$

Since lfp and gfp are dual notions, facts about one can be transformed into facts about the other by reversing the orientation of the $\subseteq$ relation and exchanging $\cap$ for $\cup$, etc.

Milner and Park's work on concurrency is an early use of coinduction. A *bisimulation* is a binary relation $r$ satisfying a property of the form $r \subseteq f\ r$. Two processes are called equivalent if the pair belongs to any bisimulation; thus, process equivalence is defined to be $\mathtt{gfp}\ f$. Two processes can be proved equivalent by exhibiting a suitable bisimulation [20].

The set LList $A$ of lazy lists over $A$ will be our main example of a greatest fixedpoint, starting in §6. The set contains both finite and infinite lists. In fact, LList $A$ and List $A$ are both fixedpoints of the same monotone function. We have List $A \subseteq$ LList $A$ and LList $A$ shares the introduction rules of List $A$, justifying the existence of finite lists.

A new principle, called *corecursion*, defines certain infinite lists; coinduction proves that these lists belong to LList $A$. Finally, the equality relation on LList $A$ happens to coincide with the gfp of a certain function. Coinduction can therefore prove equations between infinite lists. The Isabelle theory proves many familiar laws involving the append and map functions.

Coinduction can also prove that the function defined by corecursion is unique. Categorists will note that LList $A$ is a final coalgebra, just as List $A$ is an initial algebra.

# 4 Infinite trees in HOL

The lfp and gfp operators both have type $[\alpha\,\mathtt{set} \Rightarrow \alpha\,\mathtt{set}] \Rightarrow \alpha\,\mathtt{set}$. Applying them to a monotone function $f :: \tau\,\mathtt{set} \Rightarrow \tau\,\mathtt{set}$ automatically instantiates $\alpha$ to $\tau$; the result has type $\tau\,\mathtt{set}$. We should regard $\tau$ as a large space, from which lfp and gfp carve out various subspaces. It matters not if $\tau$ contains extraneous or ill-formed elements, since a suitable $f$ will discard them.

My approach is to formalize all recursive data structure definitions using one particular $\tau$, with a rich structure. Constructions — lists, trees, etc. — are sets of nodes. A node is a (*position*, *label*) pair and has type $\alpha\,\mathtt{node}$. Thus $\tau$ is $\alpha\,\mathtt{node}\,\mathtt{set}$, where $\alpha$ is the type of atoms that may occur in the construction.

Data structures are defined as sets of type $\alpha\,\mathtt{node}\,\mathtt{set}\,\mathtt{set}$. The binary operators $\otimes$ and $\oplus$, analogues of the Cartesian product and disjoint sum, take two sets of that type and yield another such set. Similarly, `List` $A$ and `LList` $A$ are unary set operators over type $\alpha$ `node set set`; they can even participate in other recursive data structure definitions.

Figure 1: An infinite tree with finite branching

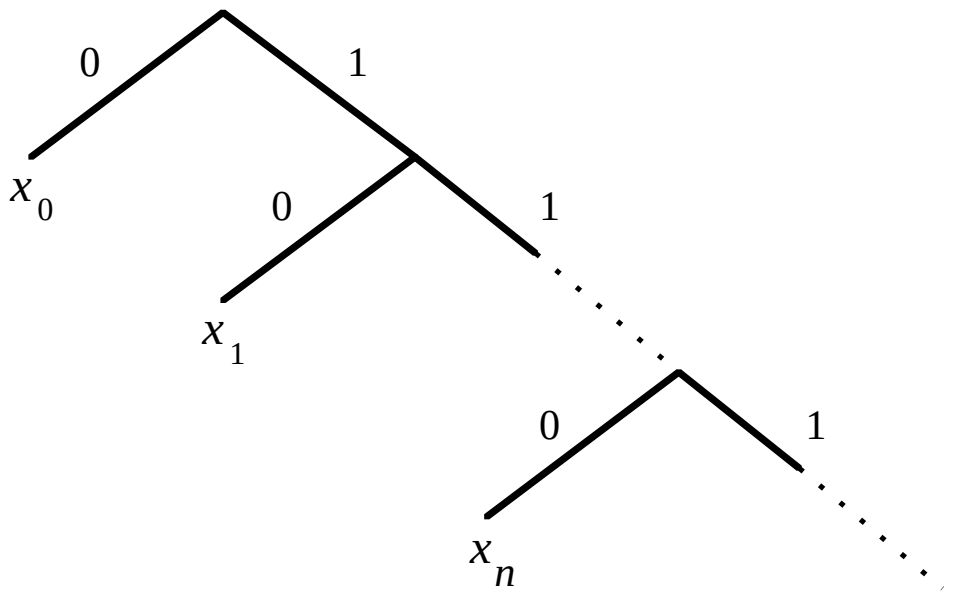

## 4.1 A possible coding of lists

To understand the definition of type $\alpha$ `node`, let us examine a simpler case, which only works for lists. The finite or infinite list $[x_0, x_1, \ldots, x_n, \ldots]$ could be represented by the set of pairs

$$\{(0, x_0), (1, x_1), \ldots, (n, x_n), \ldots\}.$$

Each pair consists of the position $m$, a natural number, paired with the label $x_m$. To 'cons' an element $a$ to the front of this list, we must add one to all the position numbers. If `succfst` $(m, x) =$ (`Suc` $m$, $x$) then `succfst` " $l$ increases all the position numbers in $l$. We may define the list constructors by

$$\begin{aligned} \mathtt{NIL} &\equiv \{\} \\ \mathtt{CONS}\ a\ l &\equiv \{(0, a)\} \cup \mathtt{succfst} \text{ `` } l \end{aligned}$$

The function $\lambda L. \{\mathtt{NIL}\} \cup (\bigcup_x \bigcup_{l \in L} \{\mathtt{CONS}\ x\ l\})$ is clearly monotone by the properties of unions. Its `lfp` is the set of finite lists; its `gfp` includes the infinite lists too. Each list has type $(\sigma \times \sigma)$ `set`, where $\sigma$ is the type of the list's elements.

## 4.2 Non-well-founded trees

In the representation above, the position of a list element is a natural number. To handle trees, let the position of a tree node be a list of natural numbers, giving the path from the root to that node. For example, the infinite tree of Figure 1 could be represented by the set of pairs

$$\{([0], x_0), ([1, 0], x_1), \ldots, (\underbrace{[1, \ldots, 1}_{n}, 0], x_n), \ldots\}.$$

The list $[1, 1, 0]$ gives the position of node $x_2$ in the tree.

Figure 2: An infinite tree with infinite branching

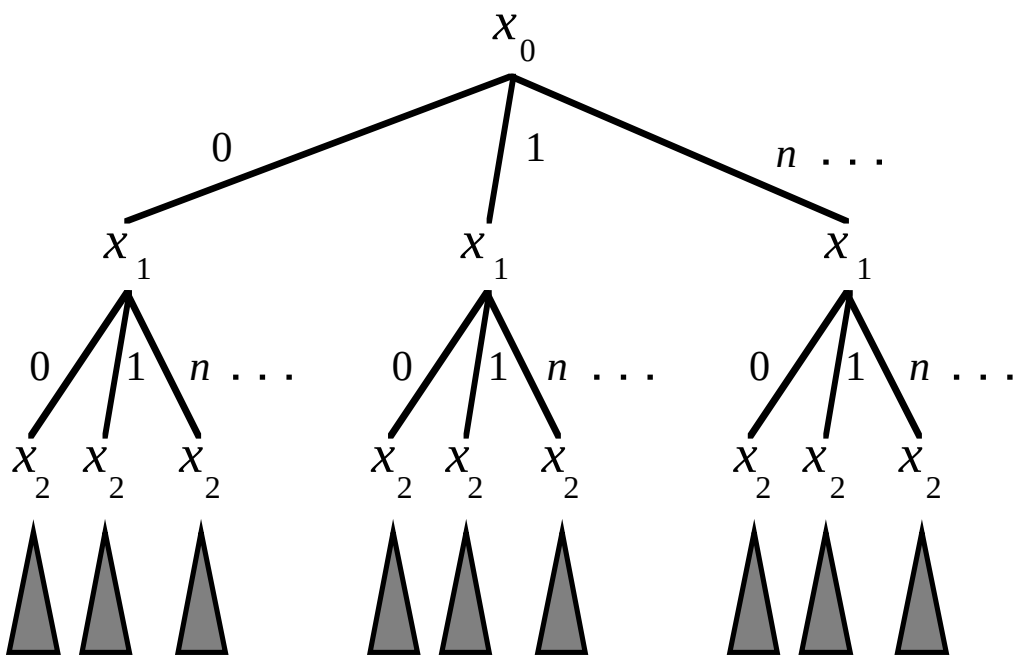

The labelled nodes determine the tree's structure. A tree that has no labelled nodes is represented by the empty set and is indistinguishable from the empty tree. Each label defines the corresponding position. Moreover, it implies the existence of all its ancestor nodes. With this representation, unlabelled branch nodes exist only because of the labelled leaf nodes underneath them.

We can even represent trees with infinite branching. The tree of Figure 2 is the infinite set of labelled nodes of the form $([k_0, k_1, \ldots, k_{n-1}], x_n)$ for $n \geq 0$ and $k_i :: \texttt{nat}$.

The definitions discussed below do not exploit the representation in its full generality. Branching is finite — binary, in fact — but the depth may be infinite.

### 4.3 The formal definition of type $\alpha\,\texttt{node}$

A node is essentially a list paired with a label. We could represent lists as described above, but this would immediately be superseded by the resulting representation of trees, a duplication of effort. Luckily, we require only lists of natural numbers. We could encode them by Gödel numbers of the form $2^{k_0}3^{k_1}\cdots$, but they are more simply represented by functions of type $\texttt{nat} \Rightarrow \texttt{nat}$.

Let the list $[k_0, \ldots, k_{n-1}]$ denote some function $f$ such that

$$f\ i = \begin{cases} \texttt{Suc}\ k_i & \text{if } 0 \leq i < n; \\ 0 & \text{if } i = n. \end{cases}$$

Note that $f\ i$ could be anything for $i > n$. The constant function $\lambda i.0$ represents the empty list. To 'cons' an element $a$ to the front of the list $f$, we use $\texttt{Push}\ a\ f$, which is defined by cases on its third argument:

$$\begin{aligned} \texttt{Push}\ a\ f\ i \equiv \texttt{case}\ i\ \texttt{of}\ & 0 \quad\ \Rightarrow \texttt{Suc}\ a \\ & \mid \texttt{Suc}\ j \Rightarrow f\ j \end{aligned}$$

Each node is represented by a pair $f\ x$, where $f :: \texttt{nat} \Rightarrow \texttt{nat}$ stands for a list and $x :: \alpha + \texttt{nat}$ is a label. The disjoint sum type allows a label to contain either an element of type $\alpha$ or a natural number; some constructions below require natural numbers in trees.

To prove that equality is a $\texttt{gfp}$, we must impose a finiteness restriction on $f$. The *take-lemma*, which says that two lazy lists are equal if all their corresponding finite initial segments are equal [5], is a standard reasoning method in lazy functional programming. The take-lemma is valid because a lazy list is nothing more than the set of its finite parts. We may similarly prove that two infinite data structures are equal, if we ensure that a tree cannot contain a node at an infinite depth. The restriction is only necessary because lists of natural numbers are represented by functions.

The set of all nodes is therefore defined as follows:

$$\texttt{Node} \equiv \{p \mid \exists f\, x\, n\, (p = (f, x) \wedge f\ n = 0)\} \qquad (2)$$

The second conjunct ensures that the position list is finite: $f\ n = 0$ for some $n$. No other conditions need to be imposed upon nodes or node sets; the fixedpoint operators exclude the undesirable elements.

Since `Node` has a complex type, namely

$$((\texttt{nat} \Rightarrow \texttt{nat}) \times (\alpha + \texttt{nat}))\, \texttt{set},$$

let us define $\alpha\, \texttt{node}$ to be the type of nodes taking labels from $\alpha$. As described in §2.5, the subtype declaration yields abstraction and representation functions:

$$\texttt{Abs_Node} :: ((\texttt{nat} \Rightarrow \texttt{nat}) \times (\alpha + \texttt{nat}))\, \texttt{set} \Rightarrow \alpha\, \texttt{node}$$
$$\texttt{Rep_Node} :: \alpha\, \texttt{node} \Rightarrow ((\texttt{nat} \Rightarrow \texttt{nat}) \times (\alpha + \texttt{nat}))\, \texttt{set}$$

## 4.4 The binary tree constructors

Possibly infinite binary trees represent all data structures in this theory. Binary trees are sets of nodes. The simplest binary tree, `Atom` $a$, consists of a label alone. If $M$ and $N$ are binary trees, then $M \cdot N$ is the tree consisting of the two branches $M$ and $N$. Thus there are two primitive tree constructors:

$$\texttt{Atom} :: \alpha + \texttt{nat} \Rightarrow \alpha\, \texttt{node}\, \texttt{set}$$
$$(\cdot) :: [\alpha\, \texttt{node}\, \texttt{set}, \alpha\, \texttt{node}\, \texttt{set}] \Rightarrow \alpha\, \texttt{node}\, \texttt{set}$$

Writing [] for $\lambda i.0$, we could define these constructors semi-formally as follows:

$$\texttt{Atom}\ a \equiv \{([], a)\}$$
$$M \cdot N \equiv \{(\texttt{Push}\ 0\ f,\ x)\}_{(f,x)\in M} \cup \{(\texttt{Push}\ 1\ f,\ x)\}_{(f,x)\in N}$$

These have the expected properties for infinite binary trees. The constructors are injective. We can recover $a$ from `Atom` $a$; we can recover $M$ and $N$ from $M \cdot N$ by stripping the initial 0 or 1. We always have `Atom` $a \neq M \cdot N$ because `Atom` $a$ contains [] as a position while $M \cdot N$ does not. Trees need not be WF; for example, there are infinite trees such that $M = M \cdot M$.

For the sake of readability, definitions below will analyse their arguments using pattern matching instead `Rep_Node`. Let us also abbreviate (`Suc` 0) as 1. The formal definitions of `Atom` and $(\cdot)$ are cumbersome, but still readable enough:

$$\texttt{Push_Node}\ k\, (\texttt{Abs_Node}(f, a)) \equiv \texttt{Abs_Node}(\texttt{Push}\ k\ f,\ a)$$
$$\texttt{Atom}\ a \equiv \{\texttt{Abs_Node}\, (\lambda i.0,\ a)\}$$
$$M \cdot N \equiv \texttt{Push_Node}\ 0 \text{ “ } M \cup \texttt{Push_Node}\ 1 \text{ “ } N$$

Now $\texttt{Push_Node}\ k$ takes the node represented by $(f, a)$ to the one represented by $(\texttt{Push}\ k\ f,\ a)$. Finally, the image $\texttt{Push_Node}\ 0$ " $M$ applies this operation upon every node in $M$.

## 4.5 Products and sums for binary trees

One objective of this theory is to justify fixedpoint definitions of the data structures, such as

$$\texttt{List}\ A \equiv \texttt{lfp}(\lambda Z.\,\{\texttt{NIL}\} \oplus (A \otimes Z)),$$

where $\otimes$ is some form of Cartesian product and $\oplus$ is some form of disjoint sum. We cannot found these upon the usual HOL ordered pairs and injections. In both $(x, y)$ and $\texttt{Inl}\ x$, the type of the result is more complex than that of the arguments; the $\texttt{lfp}$ call would be ill-typed. Therefore we must find alternative definitions of $\otimes$ and $\oplus$.

Since $(\cdot)$ is injective, we may use it as an ordered pairing operation and define the product by

$$A \otimes B \equiv \bigcup_{x\in A} \bigcup_{y\in B} \{x \cdot y\}$$

Now $A$, $B$ and $A \otimes B$ all have the same type, namely $\alpha\,\texttt{node set set}$.

Disjoint sums are typically coded in set theory by

$$A + B \equiv \{(0, x)\}_{x\in A} \cup \{(1, y)\}_{y\in B}.$$

In the present setting, this requires distinct trees to play the roles of 0 and 1. Perhaps we could find two distinct sets of nodes, such as the empty set and the universal set, but this would complicate matters below.[3] Precisely to avoid such complications, natural numbers are always allowed as labels — recall that $\texttt{Atom}$ has type $\alpha + \texttt{nat} \Rightarrow \alpha\,\texttt{node set}$. The derived constructors

$$\begin{aligned}\texttt{Leaf} &:: \alpha \Rightarrow \alpha\,\texttt{node set}\\ \texttt{Numb} &:: \texttt{nat} \Rightarrow \alpha\,\texttt{node set}\end{aligned}$$

are defined by

$$\begin{aligned}\texttt{Leaf}\ a &\equiv \texttt{Atom}(\texttt{Inl}\ a)\\ \texttt{Numb}\ k &\equiv \texttt{Atom}(\texttt{Inr}\ k)\end{aligned}$$

We may now define disjoint sums in the traditional manner:

$$\begin{aligned}\texttt{In0}\ M &\equiv \texttt{Numb}\ 0 \cdot M\\ \texttt{In1}\ N &\equiv \texttt{Numb}\ 1 \cdot N\\ A \oplus B &\equiv (\texttt{In0}\ \text{``}\ A) \cup (\texttt{In1}\ \text{``}\ B)\end{aligned}$$

Both injections have type $\alpha\,\texttt{node set} \Rightarrow \alpha\,\texttt{node set}$, while $A$, $B$ and $A \oplus B$ all have type $\alpha\,\texttt{node set set}$. Since the latter type is also closed under $\otimes$ and contains copies of

[3]The equation $\{\} \cdot \{\} = \{\}$ would frustrate the development of WF trees.

$\alpha$ and `nat`, it has enough structure to contain virtually every sort of finitely branching tree. Note that $\oplus$ and $\otimes$ are monotonic, by the monotonicity of unions and images.

The eliminators for $\otimes$ and $\oplus$ are analogous to those for the product and sum types, namely `split` and `sum_case`. They are defined in the normal manner, using descriptions, and satisfy the corresponding equations:

$$\begin{aligned} \texttt{Split}\, c\,(M \cdot N) &= c\; M\; N \\ \texttt{Case}\; c\; d\,(\texttt{In0}\; M) &= c\; M \\ \texttt{Case}\; c\; d\,(\texttt{In1}\; N) &= d\; N \end{aligned}$$

# 5 Well-founded data structures

In other work [26], I have investigated the formalization of recursive data structures in Zermelo-Frænkel (ZF) set theory. It is a general approach, allowing mutual recursion; moreover, recursive set constructions may take part in new recursive definitions, as in `term` $A = A \times$ `list`(`term` $A$). It has been mechanized within an Isabelle implementation of ZF set theory, just as the present work has been mechanized within an Isabelle implementation of HOL. It even supports non-WF data structures, using a variant form of pairing [27].

The Isabelle/HOL theory handles both WF and non-WF data structures. The WF ones are similar to those investigated in ZF, so let us dispense with them quickly.

## 5.1 Finite lists

We can now define the operator `List` to be the least solution to the recursion equation `List` $A = \{\texttt{NIL}\} \oplus (A \otimes \texttt{List}\; A)$. The Knaster-Tarski Theorem applies because $\oplus$ and $\otimes$ are monotonic. We can even prove that `List` is a monotonic operator over type $\alpha$ `node set set`, justifying definitions such as `Term` $A = A \otimes \texttt{List}(\texttt{Term}\; A)$.

The formal definition of `List` $A$ uses `lfp` to get the least fixedpoint:

$$\begin{aligned} \texttt{List_Fun}\; A &\equiv \lambda Z.\; \{\texttt{Numb}\; 0\} \oplus (A \otimes Z) \\ \texttt{List}\; A &\equiv \texttt{lfp}(\texttt{List_Fun}\; A) \\ \texttt{NIL} &\equiv \texttt{In0}(\texttt{Numb}\; 0) \\ \texttt{CONS}\; M\; N &\equiv \texttt{In1}(M \cdot N) \end{aligned}$$

From these we can easily derive the list introduction rules (§3.1), and various injectivity properties:

$$\texttt{CONS}\; M\; N \neq \texttt{NIL} \qquad (\texttt{CONS}\; K\; M = \texttt{CONS}\; L\; N) = (K = L \wedge M = N)$$

Now we can define case analysis (for recursion, see the next section). The eliminator for lists is expressed using those for $\oplus$ and $\otimes$:

$$\texttt{List_case}\; c\; d \equiv \texttt{Case}\,(\lambda x.c)\,(\texttt{Split}\; d)$$

It satisfies the expected equations:

$$\texttt{List_case}\; c\; d\; \texttt{NIL} = c \tag{3}$$

$$\texttt{List_case}\; c\; d\,(\texttt{CONS}\; M\; N) = d\; M\; N \tag{4}$$

For readability we can use this operator via Isabelle's `case` syntax.

Later, (§6), we shall define `LList` $A$, which includes infinite lists, as the greatest fixedpoint of `List_Fun` $A$. Our definitions of `NIL`, `CONS` and `List_case` will continue to work, even for the infinite lists. If $N$ is an infinite list then `CONS` $M\ N$ is also infinite.

Since `List` $A$ contains no infinite lists, we may instantiate the `lfp` induction rule to obtain structural induction (§3.1). By induction we may prove the properties expected of a WF data structure, such as

$$\frac{N \in \texttt{List}\ A}{\texttt{CONS}\ M\ N \neq N} \tag{5}$$

The Isabelle theory proceeds to define the type $\alpha\,$`list` to contain those values of type $\alpha\,$`node set` that belong to the set `List(range(Leaf))`. If $x_1, x_2, \ldots, x_n$ have type $\alpha$ then the list $[x_1, x_2, \ldots, x_n]$ is represented by

$$\texttt{CONS}\,(\texttt{Leaf}\ x_1)(\texttt{CONS}\,(\texttt{Leaf}\ x_2)(\ldots\texttt{CONS}\,(\texttt{Leaf}\ x_n)\ \texttt{NIL}\ldots)).$$

Type $\alpha\,$`list` has two constructors, `Nil` :: $\alpha\,$`list` and `Cons` :: $[\alpha, \alpha\,\texttt{list}] \Rightarrow \alpha\,\texttt{list}$, and the usual induction rule, recursion operator, etc. This type definition (§2.5) is the final stage in making lists convenient to use; the details are routine and omitted.

## 5.2 A space for well-founded types

Lisp's symbolic expressions are built up from identifiers and numbers by pairing. Formalizing S-expressions in HOL further demonstrates the Isabelle theory. More importantly, it leads to a uniform treatment of recursive functions for virtually all finitely branching WF data structures. An essential feature of S-expressions is that they are finite. Therefore, let us define them using `lfp`:

$$\texttt{Sexp} \equiv \texttt{lfp}(\lambda Z.\, \texttt{range}(\texttt{Leaf}) \cup \texttt{range}(\texttt{Numb}) \cup (Z \otimes Z)) \tag{6}$$

Observe how `range` expresses the sets of all `Leaf` $a$ and `Numb` $k$ constructions. We can develop `Sexp` in the same manner as `List` $A$, deriving an induction rule, a case analysis operator, etc.

But `Sexp` is a rather special subset of our universe, itself suitable for defining recursive data structures. It contains all constructions of the form `Leaf` $a$, `Numb` $k$ and $M \cdot N$, and therefore also `In0` $M$ and `In1` $N$. Thus it is closed under $\otimes$ and $\oplus$. Defined by `lfp`, all the constructions in it are finite.

Now `Sexp` is not necessarily large enough to contain `List` $A$ for arbitrary $A$, since $A$ might contain infinite constructions. But `Sexp` is closed under `List`: we can easily prove `List(Sexp)` $\subseteq$ `Sexp`, expressing that lists of finite constructions are themselves finite constructions. Similarly, `Sexp` is closed under many other finite data structures.

Recursion on `Sexp` gives us recursion on all these WF data structures. Isabelle's HOL theory contains a derivation of WF recursion. This justifies defining any function whose recursive calls decrease its argument under some WF relation. The *immediate subexpression* relation on `Sexp` is the set of pairs

$$(\prec) \equiv \bigcup_{M\in\texttt{Sexp}} \bigcup_{N\in\texttt{Sexp}} \{(M, M\cdot N),\ (N, M\cdot N)\}.$$

Structural induction on Sexp proves that this relation is WF. The transitive closure of a relation can be defined using lfp, and can be proved to preserve well-foundedness [26]. So $M \prec^+ N$ expresses that $M$ is a subexpression of $N$. WF recursion justifies any function on S-expressions whose recursive calls take subexpressions of the original argument.

Let us apply this to lists. Recall that

$$\texttt{CONS}\ M\ N \equiv \texttt{In1}(M \cdot N) \equiv \texttt{Numb}\ 1 \cdot (M \cdot N).$$

A sublist is therefore a subexpression. Structural recursion on lists is an instance of WF recursion. Suppose $f$ is defined by

$$\begin{aligned} f\ M \equiv \texttt{wfrec}\,(\prec^+)\ M\,(\lambda M\,g.\ \texttt{case}\ M\ \texttt{of}\ & \texttt{NIL} \Rightarrow c \\ & \mid \texttt{CONS}\ x\ y \Rightarrow d\ x\ y\,(g\ y)) \end{aligned}$$

We immediately obtain $f\ \texttt{NIL} = c$. The fact $N \prec^+ \texttt{CONS}\ M\ N$ justifies the recursion equation

$$\frac{M \in \texttt{Sexp} \qquad N \in \texttt{Sexp}}{f(\texttt{CONS}\ M\ N) = d\ M\ N\,(f\ N).}$$

Most familiar list functions — append, reverse, map — have obvious definitions by structural recursion. But the theory does not insist upon structural recursion; it can express functions such as quicksort in their natural form, using WF recursion in its full generality.

Definition (6) is not the only possible way of characterizing the set of finite constructions. We could instead formalize the finite powerset operator using lfp [26]. The set of all finite, non-empty sets of nodes would be larger than Sexp while satisfying the same key closure properties. Defining a suitable WF relation on this set might be tedious.

The Isabelle/HOL theory of WF data structures is quite general, at least for finite branching. Non-WF data structures pose greater challenges.

# 6 Lazy lists and coinduction

Defining the set of lists as a greatest instead of a least fixedpoint admits infinite as well as finite lists. We can then model computation and equality, realizing to some extent the theory of lists in lazy functional languages [5]. However, our "lazy lists" have no inbuilt operational semantics; after all, HOL can express non-computable functions.

The set of lazy lists is defined by $\texttt{LList}\ A \equiv \texttt{gfp}(\texttt{List_Fun}\ A)$; recall that List_Fun and the list operations NIL, CONS and List_case were defined in §5.1. The fixedpoint property yields introduction rules for LList $A$:

$$\texttt{NIL} \in \texttt{LList}\ A \qquad \frac{M \in A \quad N \in \texttt{LList}\ A}{\texttt{CONS}\ M\ N \in \texttt{LList}\ A}$$

These may resemble their finite list counterparts (§3.1), but they differ significantly, for CONS is well-behaved even when applied to an infinite list. In particular, the List_case equation (4) works for everything of the form CONS $M$ $N$.

Since $\texttt{LList}\ A$ is a greatest fixedpoint, it does not have a structural induction principle. Well-foundedness properties such as the list theorem (5) have no counterparts for $\texttt{LList}\ A$. We can construct a counterexample and prove that it belongs to $\texttt{LList}\ A$ by coinduction.

The weak coinduction rule for $\texttt{LList}\ A$ performs type checking for infinite lists:

$$\frac{M \in X \quad X \subseteq \texttt{List_Fun}\ A\ X}{M \in \texttt{LList}\ A} \tag{7}$$

The strong coinduction rule, as described in §3.2, implicitly includes $\texttt{LList}\ A$:

$$\frac{M \in X \quad X \subseteq \texttt{List_Fun}\ A\ X \cup \texttt{LList}\ A}{M \in \texttt{LList}\ A} \tag{8}$$

## 6.1 An infinite list

One non-WF list is the infinite list of $M$s:

$$\texttt{Lconst}\ M = \texttt{CONS}\ M\,(\texttt{Lconst}\ M) \tag{9}$$

Corecursion, a general method for defining infinite lists, is discussed below. For now, let us construct $\texttt{Lconst}\ M$ explicitly as a fixedpoint:

$$\texttt{Lconst}\ M \equiv \texttt{lfp}(\lambda Z.\,\texttt{CONS}\ M\ Z)$$

The Knaster-Tarski Theorem applies because lists are sets of nodes and CONS is monotonic in both arguments. (Recall from §4.4 that $(\cdot)$ is defined in terms of the monotonic operations union and image.) I have used lfp but gfp would work just as well — we need only the fixedpoint property (9).

We cannot prove that $\texttt{Lconst}\ M$ is a lazy list by the introduction rules alone. The coinduction rule (7) proves $\texttt{Lconst}\ M \in \texttt{LList}\{M\}$ if we can find a set $X$ containing $\texttt{Lconst}\ M$ and included in $\texttt{List_Fun}\,\{M\}\,X$. A suitable $X$ is $\{\texttt{Lconst}\ M\}$. Obviously $\texttt{Lconst}\ M \in \{\texttt{Lconst}\ M\}$. We must also show

$$\{\texttt{Lconst}\ M\} \subseteq \texttt{List_Fun}\,\{M\}\,\{\texttt{Lconst}\ M\}.$$

The fixedpoint property (9) transforms this to

$$\{\texttt{CONS}\ M\,(\texttt{Lconst}\ M)\} \subseteq \texttt{List_Fun}\,\{M\}\,\{\texttt{Lconst}\ M\},$$

which is obvious by the definitions of CONS and List_Fun.

Deriving introduction rules for List_Fun allows shorter machine proofs:

$$\texttt{NIL} \in \texttt{List_Fun}\ A\ X \tag{10}$$

$$\frac{M \in A \quad N \in X}{\texttt{CONS}\ M\ N \in \texttt{List_Fun}\ A\ X} \tag{11}$$

## 6.2 Equality of lazy lists; the take-lemma

Because HOL lazy lists are sets of nodes, the equality relation on `LList` $A$ is an instance of ordinary set equality. By investigating this relation further, we obtain nice, coinductive methods for proving that two lazy lists are equal.

Let `take` $k\ l$ return $l$'s first $k$ elements as a finite list. Bird and Wadler [5] use the take-lemma to prove equality of lazy lists $l_1$ and $l_2$:

$$\frac{\forall k\, \texttt{take}\ k\ l_1 = \texttt{take}\ k\ l_2}{l_1 = l_2}$$

It embodies a continuity principle shared by our HOL formalization.

The definition (2) of nodes (in §4.3) requires each node to have a finite depth. A node contains a pair $(f, x)$, where $x$ is the label and $f$ codes the position. Our definition ensures $f\ k = 0$ for some $k$, which is the depth of the node. We can formalize the depth directly, using the least number principle and pattern matching:

$$\begin{aligned} \texttt{LEAST}\ k.\, \phi k &\equiv \epsilon k.\, \phi k \wedge (\forall j\, (j < k \rightarrow \neg \phi j)) \\ \texttt{ndepth}\,(\texttt{Abs_Node}(f, x)) &\equiv \texttt{LEAST}\ k.\, f\ k = 0 \end{aligned}$$

Our generalization of `take`, called `ntrunc`, applies to all data structures, not just lists. It returns the set of all nodes having less than a given depth:

$$\texttt{ntrunc}\ k\ N \equiv \{nd \mid nd \in N \wedge \texttt{ndepth}\ nd < k\}$$

Elementary reasoning derives results describing `ntrunc`'s effect upon various constructions:

$$\begin{aligned} \texttt{ntrunc}\ 0\ M &= \{\} \\ \texttt{ntrunc}\,(\texttt{Suc}\ k)\,(\texttt{Leaf}\ a) &= \texttt{Leaf}\ a \\ \texttt{ntrunc}\,(\texttt{Suc}\ k)\,(\texttt{Numb}\ a) &= \texttt{Numb}\ a \\ \texttt{ntrunc}\,(\texttt{Suc}\ k)\,(M \cdot N) &= \texttt{ntrunc}\ k\ M \cdot \texttt{ntrunc}\ k\ N \end{aligned}$$

Since $\texttt{In0}\ M \equiv \texttt{Numb}\ 0 \cdot M$ we obtain

$$\begin{aligned} \texttt{ntrunc}\ 1\,(\texttt{In0}\ M) &= \{\} \\ \texttt{ntrunc}\,(\texttt{Suc}(\texttt{Suc}\ k))\,(\texttt{In0}\ M) &= \texttt{In0}(\texttt{ntrunc}\,(\texttt{Suc}\ k)\ M) \end{aligned}$$

and similarly for `In1`.

Our generalization of `take` enjoys a generalization of the take-lemma:

**Lemma 1** If $\texttt{ntrunc}\ k\ M = \texttt{ntrunc}\ k\ N$ for all $k$ then $M = N$.

This obvious fact is a key result. It gives us a method for proving the equality of any constructions $M$ and $N$. We could apply this "ntrunc-lemma" directly, but instead we shall package it into a form suitable for coinduction.

## 6.3 Diagonal set operators

In order to prove list equations by coinduction, we must demonstrate that the equality relation is the greatest fixedpoint of some monotone operator. To this end, we define diagonal set operators for $\otimes$ and $\oplus$. A *diagonal set* has the form $\{(x,x)\}_{x\in A}$, internalizing the equality relation on $A$.

A binary relation on sets of nodes has type $(\alpha\,\texttt{node}\,\texttt{set} \times \alpha\,\texttt{node}\,\texttt{set})\,\texttt{set}$. The operators $\otimes_D$ and $\oplus_D$ combine two such relations to yield a third. The operator `diag`, of type $\alpha\,\texttt{set} \Rightarrow (\alpha \times \alpha)\,\texttt{set}$, constructs arbitrary diagonal sets.

$$\begin{aligned}
\texttt{diag}\, A &\equiv \bigcup\nolimits_{x\in A} \{(x,x)\} \\
r \otimes_D s &\equiv \bigcup\nolimits_{(x,x')\in r} \bigcup\nolimits_{(y,y')\in s} \{(x\cdot y,\, x'\cdot y')\} \\
r \oplus_D s &\equiv \bigcup\nolimits_{(x,x')\in r} \{(\texttt{In0}\, x,\, \texttt{In0}\, x')\} \cup \bigcup\nolimits_{(y,y')\in s} \{(\texttt{In1}\, y,\, \texttt{In1}\, y')\}
\end{aligned}$$

These enjoy readable introduction rules. For $\otimes_D$ we have

$$\frac{(M,M')\in r \qquad (N,N')\in s}{(M\cdot N,\, M'\cdot N')\in r\otimes_D s}$$

while for $\oplus_D$ we have the pair of rules

$$\frac{(M,M')\in r}{(\texttt{In0}\, M,\, \texttt{In0}\, M')\in r\oplus_D s} \qquad \frac{(N,N')\in s}{(\texttt{In1}\, N,\, \texttt{In1}\, N')\in r\oplus_D s}$$

The idea is that $\otimes_D$ and $\oplus_D$ build relations in the same manner as $\otimes$ and $\oplus$ build sets. Since $\texttt{fst}$ “ $r$ is the first projection of the relation $r$, we can summarize the idea by three obvious equations:

$$\begin{aligned}
\texttt{fst}\text{ “ }\texttt{diag}\, A &= A \\
\texttt{fst}\text{ “ }(r\otimes_D s) &= (\texttt{fst}\text{ “ }r)\otimes(\texttt{fst}\text{ “ }s) \\
\texttt{fst}\text{ “ }(r\oplus_D s) &= (\texttt{fst}\text{ “ }r)\oplus(\texttt{fst}\text{ “ }s)
\end{aligned}$$

Category theorists may note that $\otimes$ and $\oplus$ are functors on a category of sets where the morphisms are binary relations; in this category, $\otimes_D$ and $\oplus_D$ give the functors' effects on the morphisms. Next, we shall do the same thing to the functor `LList`.

## 6.4 Equality of lazy lists as a `gfp`

Just as $\otimes_D$ and $\oplus_D$ extend $\otimes$ and $\oplus$ to act upon relations, let `LListD` extend `LList`. Thanks to our new operators, the definition is simple and resembles that of `LList`:

$$\begin{aligned}
\texttt{LListD_Fun}\, r &\equiv \lambda Z.\, \texttt{diag}\{\texttt{Numb}\, 0\} \oplus_D (r\otimes_D Z) \\
\texttt{LListD}\, r &\equiv \texttt{gfp}(\texttt{LListD_Fun}\, r)
\end{aligned}$$

The theorem that list equality is a `gfp` can now be stated as a succinct equation between relations: $\texttt{LListD}(\texttt{diag}\ A) = \texttt{diag}(\texttt{LList}\ A)$. Here $\texttt{diag}(\texttt{LList}\ A)$ is the equality relation on $\texttt{LList}\ A$, while $\texttt{LListD}(\texttt{diag}\ A)$ is the `gfp` of $\texttt{LListD_Fun}(\texttt{diag}\, A)$. Proving this requires another lemma about `ntrunc`.

**Lemma 2** $\forall M\, N\, [(M, N) \in \texttt{LListD}(\texttt{diag}\ A) \rightarrow \texttt{ntrunc}\ k\ M = \texttt{ntrunc}\ k\ N]$.

**Proof** By complete induction on $k$, we may assume the formula above after replacing $k$ by any smaller natural number $j$. By the fixedpoint property

$$\texttt{LListD}(\texttt{diag}\ A) = \texttt{diag}\{\texttt{Numb}\ 0\} \oplus_D (r \otimes_D \texttt{LListD}(\texttt{diag}\ A)),$$

if $(M, N) \in \texttt{LListD}(\texttt{diag}\ A)$ then there are two cases. If

$$M = N = \texttt{In0}(\texttt{Numb}\ 0) = \texttt{NIL}$$

then $\texttt{ntrunc}\ k\ M = \texttt{ntrunc}\ k\ N$ is trivial. Otherwise $M = \texttt{CONS}\ x\ M'$ and $N = \texttt{CONS}\ x\ N'$, where $(M', N') \in \texttt{LListD}(\texttt{diag}\ A)$. Recall the definition $\texttt{CONS}\ x\ y \equiv \texttt{In1}(x \cdot y)$ and the properties of $\texttt{ntrunc}$ (§6.2); we obtain

$$\texttt{ntrunc}\ k\,(\texttt{CONS}\ x\ y) = \begin{cases} \{\} & \text{if } k < 2\text{, and} \\ \texttt{CONS}\,(\texttt{ntrunc}\ j\ x)\,(\texttt{ntrunc}\ j\ y) & \text{if } k = \texttt{Suc}(\texttt{Suc}\ j). \end{cases}$$

If $k = \texttt{Suc}(\texttt{Suc}\ j)$ then $\texttt{ntrunc}\ k\ M = \texttt{ntrunc}\ k\ N$ reduces to an instance of the induction hypothesis, $\texttt{ntrunc}\ j\ M' = \texttt{ntrunc}\ j\ N'$.

Now we can prove that equation.

**Proposition 3** $\texttt{LListD}(\texttt{diag}\ A) = \texttt{diag}(\texttt{LList}\ A)$.

**Proof** Combining Lemmas 1 and 2 yields half of our desired result, $\texttt{LListD}(\texttt{diag}\ A) \subseteq \texttt{diag}(\texttt{LList}\ A)$. This is the more important half: it lets us show $M = N$ by showing $(M, N) \in \texttt{LListD}(\texttt{diag}\ A)$, which can be done using coinduction.

The opposite inclusion, $\texttt{diag}(\texttt{LList}\ A) \subseteq \texttt{LListD}(\texttt{diag}\ A)$, follows by showing that $\texttt{diag}(\texttt{LList}\ A)$ is a fixedpoint of $\texttt{LListD_Fun}(\texttt{diag}\,A)$, since $\texttt{LListD}(\texttt{diag}\ A)$ is the greatest fixedpoint. This argument is an example of coinduction.

## 6.5 Proving lazy list equality by coinduction

The weak coinduction rule for list equality yields $M = N$ provided $(M, N) \in r$ where $r$ is a suitable bisimulation between lazy lists:

$$\frac{(M, N) \in r \qquad r \subseteq \texttt{LListD_Fun}(\texttt{diag}\,A)r}{M = N} \tag{12}$$

Coinduction has many variant forms (§3.2). Strong coinduction includes the equality relation implicitly in every bisimulation:

$$\frac{(M, N) \in r \qquad r \subseteq \texttt{LListD_Fun}(\texttt{diag}\,A)r \cup \texttt{diag}(\texttt{LList}\ A)}{M = N} \tag{13}$$

Expanding the definitions of NIL, CONS and LListD_Fun creates unwieldy formulae. The Isabelle theory derives two rules to avoid this, resembling the List_Fun rules (10) and (11):

$$(\texttt{NIL}, \texttt{NIL}) \in \texttt{LListD_Fun}(\texttt{diag}\,A)r \tag{14}$$

$$\frac{x \in A \quad (M, N) \in r}{(\texttt{CONS}\ x\ M, \texttt{CONS}\ x\ N) \in \texttt{LListD_Fun}(\texttt{diag}\,A)r} \tag{15}$$

# 7 Lazy lists and corecursion

We have defined the infinite list $\texttt{Lconst}\ M = [M, M, \ldots]$ using a fixedpoint. The construction clearly generalizes to other repetitive lists such as $[M, N, M, N, \ldots]$. But how can we define infinite lists such as $[1, 2, 3, \ldots]$? And how can we define the usual list operations, like append and map? Structural recursive definitions would work for elements of $\texttt{List}\ A$ but not for the infinite lists in $\texttt{LList}\ A$.

Corecursion is a dual form of structural recursion. Recursion defines functions that consume lists, while corecursion defines functions that create lists. Corecursion originated in the category theoretic notion of final coalgebra; Mendler [19] and Geuvers [12], among others, have investigated it in type theories.

This paper does not attempt to treat corecursion categorically. And instead of describing the general case in all its complexity, it simply treats a key example: lazy lists. Let us begin with motivation and examples.

## 7.1 Introduction to corecursion

Corecursion defines a lazy list in terms of some seed value $a :: \alpha$ and a function $f :: \alpha \Rightarrow \texttt{unit} + (\beta\,\texttt{node}\,\texttt{set} \times \alpha)$. Recall from §2.3 that $\texttt{unit}$ is the nullary product type, whose sole value is $()$, while $\times$ and $+$ are the product and sum type operators. Thus $\texttt{LList_corec}$ has type

$$[\alpha, \alpha \Rightarrow \texttt{unit} + (\beta\,\texttt{node}\,\texttt{set} \times \alpha)] \Rightarrow \beta\,\texttt{node}\,\texttt{set}.$$

It must satisfy

$$\texttt{LList_corec}\ a\ f = \begin{cases} \texttt{NIL} & \text{if } f\ a = \texttt{Inl}\,(); \\ \texttt{CONS}\ x\,(\texttt{LList_corec}\ b\ f) & \text{if } f\ a = \texttt{Inr}\,(x, b). \end{cases}$$

The idea should be clear: $f$ takes the seed $a$ and either returns $\texttt{Inl}\,()$, to end the list here, or returns $\texttt{Inr}\,(x, b)$, to continue the list with next element $x$ and seed $b$. By keeping the seed forever $M$ and always returning it as the next element, corecursion can express $\texttt{Lconst}\ M$:

$$\texttt{Lconst}\ M \equiv \texttt{LList_corec}\ M(\lambda N.\texttt{Inr}\,(N, N))$$

Consider the functional $\texttt{Lmap}$, which applies a function to every element of a list:

$$\texttt{Lmap}\ g\ [x_0, x_1, \ldots, x_n, \ldots] = [g\,x_0, g\,x_1, \ldots, g\,x_n, \ldots]$$

The usual recursion equations are

$$\texttt{Lmap}\ g\,\texttt{NIL} = \texttt{NIL} \tag{16}$$

$$\texttt{Lmap}\ g\,(\texttt{CONS}\ M\ N) = \texttt{CONS}\,(g\ M)\,(\texttt{Lmap}\ g\ N). \tag{17}$$

Corecursion handles these easily. To compute $\texttt{Lmap}\ g\ M$, take $M$ as the seed. If $M = \texttt{NIL}$ then end the result list; if $M = \texttt{CONS}\ x\ M'$ then continue the result list with next element $g\ x$ and seed $M'$. The formal definition uses $\texttt{List_case}$ to inspect $M$:

$$\begin{aligned} \texttt{Lmap}\ g\ M \equiv \texttt{LList_corec}\ M\,(\lambda M.\ \texttt{case}\ M\ \texttt{of}\ & \texttt{NIL} \Rightarrow \texttt{Inl}\,() \\ & \mid \texttt{CONS}\ x\ M' \Rightarrow \texttt{Inr}\,(g\ x,\ M')) \end{aligned}$$

This definition of map has little in common with the standard recursive one. Append comes out stranger still, and other standard functions seem to be lost altogether.

## 7.2 Harder cases for corecursion

With corecursion, the case analysis is driven by the output list, rather than the input list. The append function highlights this peculiarity. The usual recursion equations for append perform case analysis on the first argument:

$$\begin{aligned} \texttt{Lappend}\,\texttt{NIL}\;N &= N \\ \texttt{Lappend}\,(\texttt{CONS}\;M_1\;M_2)\;N &= \texttt{CONS}\;M_1\,(\texttt{Lappend}\;M_2\;N) \end{aligned}$$

But a `NIL` input does not guarantee a `NIL` output, as it did for `Lmap`; consider

$$\texttt{Lappend}\,\texttt{NIL}\,(\texttt{CONS}\;M\;N) = \texttt{CONS}\;M\;N.$$

The correct equations for corecursion involve both arguments:

$$\begin{aligned} \texttt{Lappend}\,\texttt{NIL}\,\texttt{NIL} &= \texttt{NIL} && (18) \\ \texttt{Lappend}\,\texttt{NIL}\,(\texttt{CONS}\;N_1\;N_2) &= \texttt{CONS}\;N_1\,(\texttt{Lappend}\,\texttt{NIL}\;N_2) && (19) \\ \texttt{Lappend}\,(\texttt{CONS}\;M_1\;M_2)\;N &= \texttt{CONS}\;M_1\,(\texttt{Lappend}\;M_2\;N) && (20) \end{aligned}$$

The second line above forces `Lappend NIL` $N$ to continue executing until it has made a copy of $N$. This looks inefficient. In the context of the polymorphic $\lambda$-calculus, Geuvers [12] discusses stronger forms of corecursion that allow the seed to return an entire list at once. But my HOL theory has no operational significance; efficiency is meaningless; we may as well keep corecursion simple.[4]

The seed for `Lappend` $M\;N$ is the pair $(M, N)$. The corecursive definition performs case analysis on both lists:

- If $M = \texttt{NIL}$ then it looks at $N$:
  - If $N = \texttt{NIL}$ then end the result list.
  - If $N = \texttt{CONS}\;N_1\;N_2$ then continue the result list with next element $N_1$ and seed $(\texttt{NIL}, N_2)$.
- If $M = \texttt{CONS}\;M_1\;M_2$ then continue the result list with next element $M_1$ and seed $(M_2, N)$.

We can formalize this using `LList_corec`. Define $f$ by case analysis:

$$\begin{aligned} f \equiv \lambda(M,N).\;&\texttt{case}\;M\;\texttt{of} \\ &\qquad \texttt{NIL} \Rightarrow (\texttt{case}\;N\;\texttt{of}\;\;\texttt{NIL} \Rightarrow \texttt{Inl}\,() \\ &\qquad\qquad\qquad\qquad\qquad\;\; \mid \texttt{CONS}\;N_1\;N_2 \Rightarrow \texttt{Inr}\,(N_1, (\texttt{NIL}, N_2))) \\ &\qquad \mid \texttt{CONS}\;M_1\;M_2 \Rightarrow \texttt{Inr}\,(M_1, (M_2, N)) \end{aligned}$$

Then put `Lappend` $M\;N \equiv$ `LList_corec` $(M, N)\;f$.

[4] This relates to difficulties with formalizing (co)inductive definitions in type theory. Under some formalizations, the constructors of coinductive types are complicated and inefficient — and dually, so are the destructors of inductive types. This does not occur in my HOL treatment: constructors and destructors are directly available from the fixedpoint equations. Thus we must not push the analogy with type theory too far.

**Concatenation.** Now let us try to define a function to flatten a list of lists according to the following recursion equations:

$$\texttt{Lflat}\,\texttt{NIL} = \texttt{NIL} \tag{21}$$
$$\texttt{Lflat}(\texttt{CONS}\;M\;N) = \texttt{Lappend}\;M\;(\texttt{Lflat}\;N) \tag{22}$$

Since corecursion is driven by the output, we need to know when a CONS is produced. We can refine equation (22) by further case analysis:

$$\texttt{Lflat}(\texttt{CONS}\,\texttt{NIL}\;N) = \texttt{Lflat}\;N$$
$$\texttt{Lflat}(\texttt{CONS}\,(\texttt{CONS}\;M_1\;M_2)\;N) = \texttt{CONS}\;M_1\;(\texttt{Lflat}(\texttt{CONS}\;M_2\;N))$$

Unfortunately, the outcome of CONS NIL $N$ is still undecided; it could be NIL or CONS. Given the argument Lconst NIL — an infinite list of NILs — Lflat should run forever. There is no effective way to check whether an infinite list contains a non-NIL element.

I do not know of any natural formalization of Lflat. Since HOL can formalize non-computable functions, we can force Lflat (Lconst NIL) = NIL using a description. Such a definition would be of little relevance to computer science. Really Lflat (Lconst NIL) should be undefined, but HOL does not admit partial functions.

The filter functional, which removes from a list all elements that fail to satisfy a given predicate, poses similar problems. If no list elements satisfy the predicate, then the result is logically NIL, but there is no effective way to reach this result. Leclerc and Paulin-Mohring [17] discuss approaches to this problem in the context of the Coq system.

## 7.3 Deriving corecursion

The characteristic equation for corecursion can be stated using case analysis and pattern matching:

$$\begin{aligned}\texttt{LList_corec}\;a\;f = \texttt{case}\;f\;a\;\texttt{of} & \\ \texttt{Inl}\;u \quad &\Rightarrow \texttt{NIL} \\ \mid \texttt{Inr}(x,b) &\Rightarrow \texttt{CONS}\;x\;(\texttt{LList_corec}\;b\;f)\end{aligned} \tag{23}$$

To realize this equation, recall that an element of LList $A$ is a set of nodes, each of finite depth. A primitive recursive function can approximate the corecursion operator for nodes up to some given depth $k$:

$$\begin{aligned}\texttt{lcorf}\;0\;a\;f &= \{\} \\ \texttt{lcorf}\;(\texttt{Suc}\;k)\;a\;f &= \texttt{case}\;f\;a\;\texttt{of} \\ &\quad\;\; \texttt{Inl}\;u \quad \Rightarrow \texttt{NIL} \\ &\quad \mid \texttt{Inr}(x,b) \Rightarrow \texttt{CONS}\;x\;(\texttt{lcorf}\;k\;b\;f)\end{aligned}$$

Now we can easily define corecursion:

$$\texttt{LList_corec}\;a\;f \equiv \bigcup_k \texttt{lcorf}\;k\;a\;f$$

Equation (23) can then be proved as two separate inclusions, with simple reasoning about unions and (in one direction) induction on $k$.

The careful reader may have noticed that lcorf $k$ $a$ $f$ is not strictly primitive recursive, because the parameter $a$ varies in the recursive calls. To be completely formal, we must define a function over $a$ using higher-order primitive recursion. In the Isabelle theory, lcorf $k$ $a$ $f$ becomes LList_corec_fun $k$ $f$ $a$. It can be defined by primitive recursion:

$$\begin{aligned} \texttt{LList_corec_fun}\ 0\ f &= \lambda a.\,\{\} \\ \texttt{LList_corec_fun}\ (\texttt{Suc}\ k)\ f &= \lambda a.\,\texttt{case}\ f\ a\ \texttt{of} \\ &\qquad\quad \texttt{Inl}\ u \quad\ \ \Rightarrow \texttt{NIL} \\ &\qquad\ \mid \texttt{Inr}(x,b) \Rightarrow \texttt{CONS}\ x\ (\texttt{LList_corec_fun}\ k\ f\ b) \end{aligned}$$

Obscure as this may look, it trivially implies the equations for lcorf given above.

# 8 Examples of coinduction and corecursion

This section demonstrates defining functions by corecursion and verifying their equational and typing properties by coinduction. All these proofs have been checked by machine.

## 8.1 A type-checking rule for LList_corec

One advantage of LList_corec over ad-hoc definitions is that the result is certain to be a lazy list. To express this well-typing property in the simplest possible form, let $U :: \alpha\, \texttt{node set set}$ abbreviate the universal set of constructions: $U \equiv \{x \mid \texttt{True}\}$. We can now state a typing result:

**Example 4** LList_corec $a$ $f \in$ LList $U$.

**Proof** Apply the weak coinduction rule (7) to the set

$$V \equiv \texttt{range}(\lambda x.\, \texttt{LList_corec}\ x\ f).$$

We must show $V \subseteq \texttt{List_Fun}\ U\ V$, which reduces to

$$\texttt{LList_corec}\ a\ f \in \texttt{List_Fun}\ U\ V.$$

There are two cases. We simplify each case using equation (23):

- If $f\ a = \texttt{Inl}\ ()$, it reduces to $\texttt{NIL} \in \texttt{List_Fun}\ U\ V$, which holds by Rule (10).
- If $f\ a = \texttt{Inr}\,(x, b)$, it reduces to

  $$\texttt{CONS}\ x\ (\texttt{LList_corec}\ b\ f) \in \texttt{List_Fun}\ U\ V.$$

  This holds by Rule (11) because $x \in U$ and $\texttt{LList_corec}\ b\ f \in V$.

This result, referring to the universal set, may seem weak compared with our previous well-typing result, $\texttt{Lconst}\ M \in \texttt{LList}\{M\}$, from §6.1. It is difficult to bound the set of possible values for $x$ for the case $f\ a = \texttt{Inr}\,(x, b)$. Sharper results can be obtained by performing separate coinduction proofs for each function defined by corecursion.

## 8.2 The uniqueness of corecursive functions

Equation (23) characterizes LList_corec uniquely; the proof is a typical example of proving an equation by coinduction. Let us define an abbreviation for the corecursion property:

$$\texttt{is_corec}\ f\ h \equiv \forall u\,[h\ u = \texttt{case}\ f\ u\ \texttt{of}\ \ \texttt{Inl}\ v \quad \Rightarrow \texttt{NIL} \\ \mid \texttt{Inr}(x,b) \Rightarrow \texttt{CONS}\ x\,(h\ b)]$$

Existence, namely $\texttt{is_corec}\ f\,(\lambda x.\,\texttt{LList_corec}\ x\ f)$, is immediate by equation (23). Let us now prove uniqueness.

**Proposition 5** If $\texttt{is_corec}\ f\ h_1$ and $\texttt{is_corec}\ f\ h_2$ then $h_1 = h_2$.

**Proof** By extensionality it suffices to prove $h_1\ u = h_2\ u$ for all $u$. Now consider the bisimulation $\{(h_1\ u, h_2\ u)\}_u$, formalized in HOL by

$$r \equiv \texttt{range}(\lambda u.\,(h_1\ u, h_2\ u)).$$

Apply the coinduction rule (12), with $r$ as above and putting $A \equiv U$, the universal set. We must show $r \subseteq \texttt{LListD_Fun}(\texttt{diag}\,U)r$; since elements of $r$ have the form $(h_1\ u, h_2\ u)$ for arbitrary $u$, it suffices to show

$$(h_1\ u, h_2\ u) \in \texttt{LListD_Fun}(\texttt{diag}\,U)r.$$

There are two cases, determined by $f(u)$. In each, we apply the corecursion properties of $h_1$ and $h_2$:

- If $f\ u = \texttt{Inl}\,()$, it reduces to $(\texttt{NIL},\texttt{NIL}) \in \texttt{LListD_Fun}(\texttt{diag}\,U)r$, which holds by Rule (14).
- If $f\ u = \texttt{Inr}\,(x,b)$, it reduces to

  $$(\texttt{CONS}\ x\,(h_1\ b), \texttt{CONS}\ x\,(h_2\ b)) \in \texttt{LListD_Fun}(\texttt{diag}\,U)r.$$

  This holds by Rule (15) because $x \in U$ and $(h_1\ b, h_2\ b) \in r$.

Note the similarity between this coinduction proof and the previous one, even though the former proves set membership while the latter proves an equation. The role of $r$ in this equality proof is reminiscent of process equivalence proofs in CCS [20], where the bisimulation associates corresponding states of two processes. The equality can also be proved using Lemma 1 directly, by complete induction on $k$; the proof is considerably more complex.

## 8.3 A proof about the map functional

To demonstrate that the theory has some relevance to lazy functional programming, this section proves a simple result: that map distributes over composition.

**Example 6** If $M \in \texttt{LList}\ A$ then $\texttt{Lmap}\,(f \circ g)\ M = \texttt{Lmap}\ f\,(\texttt{Lmap}\ g\ M)$.

**Proof** The bisimulation $\{(\texttt{Lmap}\,(f \circ g)\; u,\; \texttt{Lmap}\; f\,(\texttt{Lmap}\; g\; u))\}_{u \in \texttt{LList}\; A}$ may be formalized using the image operator:

$$r \equiv (\lambda u.\,(\texttt{Lmap}\,(f \circ g)\; u,\; \texttt{Lmap}\; f\,(\texttt{Lmap}\; g\; u)))\;``\,\texttt{LList}\; A$$

As in the previous coinduction proof, apply Rule (12). The key is to show

$$(\texttt{Lmap}\,(f \circ g)\; u,\; \texttt{Lmap}\; f\,(\texttt{Lmap}\; g\; u)) \in \texttt{LListD_Fun}(\texttt{diag}\, U)r$$

for arbitrary $u \in \texttt{LList}\; A$. There are again two cases, this time depending on the form of $u$. We simplify each case using the recursion equations for `Lmap`, which follow from its corecursive definition (§7.1).

- If $u = \texttt{NIL}$, the goal reduces to $(\texttt{NIL}, \texttt{NIL}) \in \texttt{LListD_Fun}(\texttt{diag}\, U)r$, which holds by Rule (14).
- If $u = \texttt{CONS}\; M'\; N$, it reduces to

$$\begin{array}{l}(\texttt{CONS}\,(f\,(g\; M'))\,(\texttt{Lmap}\,(f \circ g)\; N), \\ \;\texttt{CONS}\,(f\,(g\; M'))\,(\texttt{Lmap}\; f\,(\texttt{Lmap}\; g\; N))) \;\in\; \texttt{LListD_Fun}(\texttt{diag}\, U)r.\end{array}$$

  This holds by Rule (15) because $f\,(g\; M') \in U$ and

$$(\texttt{Lmap}\,(f \circ g)\; N,\; \texttt{Lmap}\; f\,(\texttt{Lmap}\; g\; N)) \in r.$$

The coinductive argument bears little resemblance to the usual proof in the Logic for Computable Functions (LCF) [22, page 283]. On the other hand, all these coinductive proofs have a monotonous regularity. A similar argument proves $\texttt{Lmap}\,(\lambda x.x)\; M = M$.

## 8.4 Proofs about the list of iterates

Consider the function `Iterates` defined by

$$\texttt{Iterates}\; f\; M \equiv \texttt{LList_corec}\; M(\lambda M.\,\texttt{Inr}\,(M, f\; M)).$$

A generalization of `Lconst`, it constructs the infinite list $[M, f\,M, \ldots, f^n\,M, \ldots]$ by the recursion equation

$$\texttt{Iterates}\; f\; M = \texttt{CONS}\; M\,(\texttt{Iterates}\; f\,(f\; M)).$$

The equation

$$\texttt{Lmap}\; f\,(\texttt{Iterates}\; f\; M) = \texttt{Iterates}\; f\,(f\; M) \tag{24}$$

has a straightforward coinductive proof, using the bisimulation

$$\{(\texttt{Lmap}\; f\,(\texttt{Iterates}\; f\; u),\; \texttt{Iterates}\; f\,(f\; u))\}_{u \in \texttt{LList}\; A}.$$

Combining the previous two equations yields a new recursion equation, involving `Lmap`:

$$\texttt{Iterates}\; f\; M = \texttt{CONS}\; M\,(\texttt{Lmap}\; f\,(\texttt{Iterates}\; f\; M))$$

Harder is to show that this equation uniquely characterizes `Iterates` $f$.

**Example 7** If $h\ x = \texttt{CONS}\ x\ (\texttt{Lmap}\ f\ (h\ x))$ for all $x$ then $h = \texttt{Iterates}\ f$.

**Proof** The bisimulation for this proof is unusually complex. Let $f^n$ stand for the function that applies $f$ to its argument $n$ times. Since Lmap is a curried function, the function $(\texttt{Lmap}\ f)^n$ applies $\texttt{Lmap}\ f$ to a given list $n$ times. The bisimulation has two index variables:

$$r \equiv \{((\texttt{Lmap}\ f)^n\ (h\ u),\ (\texttt{Lmap}\ f)^n\ (\texttt{Iterates}\ f\ u))\}_{u \in \texttt{LList}\, U,\ n \geq 0}$$

Recall that $U$ stands for the universal set of the appropriate type.

Then note two facts, both easily justified by induction on $n$:

$$(\texttt{Lmap}\ f)^n\ (\texttt{CONS}\ b\ M) = \texttt{CONS}\ (f^n\ b)\ ((\texttt{Lmap}\ f)^n\ M) \tag{25}$$

$$f^n\ (f\ x) = f^{\texttt{Suc}\ n}\ x \tag{26}$$

By extensionality it suffices to prove $h\ u = \texttt{Iterates}\ f\ u$ for all $u$. Again we apply the weak coinduction rule (12), with the bisimulation $r$ shown above. The key step in verifying the bisimulation is to show

$$((\texttt{Lmap}\ f)^n\ (h\ u),\ (\texttt{Lmap}\ f)^n\ (\texttt{Iterates}\ f\ u)) \in \texttt{LListD_Fun}(\texttt{diag}\, U) r$$

for arbitrary $n \geq 0$ and $u \in \texttt{LList}\ U$. By the recursion equations for $h$ and Iterates, the pair expands to

$$\begin{array}{l}((\texttt{Lmap}\ f)^n\ (\texttt{CONS}\ u\ (\texttt{Lmap}\ f\ (h\ u))),\\ \ (\texttt{Lmap}\ f)^n\ (\texttt{CONS}\ u\ (\texttt{Iterates}\ f\ (f\ u))))\end{array}$$

and now two applications of equation (25) yield

$$\begin{array}{l}(\texttt{CONS}\ (f^n\ u)\ ((\texttt{Lmap}\ f)^n\ (\texttt{Lmap}\ f\ (h\ u))),\\ \ \texttt{CONS}\ (f^n\ u)\ ((\texttt{Lmap}\ f)^n\ (\texttt{Iterates}\ f\ (f\ u))))\end{array}$$

Applying Rule (15) to show membership in $\texttt{LListD_Fun}(\texttt{diag}\, U) r$, we are left with the subgoal

$$((\texttt{Lmap}\ f)^n\ (\texttt{Lmap}\ f\ (h\ u)),\ (\texttt{Lmap}\ f)^n\ (\texttt{Iterates}\ f\ (f\ u))) \in r.$$

By equations (24) and (26) we obtain

$$(\texttt{Lmap}\ f^{\texttt{Suc}\ n}\ (h\ u),\ \texttt{Lmap}\ f^{\texttt{Suc}\ n}\ (\texttt{Iterates}\ f\ u)) \in r,$$

which is obviously true, by the definition of $r$.

Pitts [28] describes a similar coinduction proof in domain theory. The function Iterates does not lend itself to reasoning by structural induction, but is amenable to Scott's fixedpoint induction; I proved equation (24) in the Logic for Computable Functions (LCF) [22, page 286].

## 8.5 Reasoning about the append function

Many accounts of structural induction start with proofs about append. But append is not an easy example for coinduction. Remember that the corecursive definition does not give us the usual pair of equations, but rather the three equations (18)–(20). Proofs by the weak coinduction rule (12) typically require the same three-way case analysis. Consider proving that map distributes over append:

**Example 8** If $M \in \texttt{LList}\ A$ and $N \in \texttt{LList}\ A$ then

$$\texttt{Lmap}\ f\,(\texttt{Lappend}\ M\ N) = \texttt{Lappend}\,(\texttt{Lmap}\ f\ M)\,(\texttt{Lmap}\ f\ N).$$

**Proof** Applying the weak coinduction rule with the bisimulation

$$\{(\texttt{Lmap}\ f\,(\texttt{Lappend}\ u\ v),\ \texttt{Lappend}\,(\texttt{Lmap}\ f\ u)\,(\texttt{Lmap}\ f\ v))\}_{u\in\texttt{LList}\ A, v\in\texttt{LList}\ A},$$

we must show

$$\begin{array}{l}(\texttt{Lmap}\ f\,(\texttt{Lappend}\ u\ v),\\ \ \texttt{Lappend}\,(\texttt{Lmap}\ f\ u)\,(\texttt{Lmap}\ f\ v)) \ \in\ \texttt{LListD_Fun}(\texttt{diag}\,U)r.\end{array}$$

Considering the form of $u$ and $v$, there are three cases:

- If $u = v = \texttt{NIL}$, the pair reduces by equations (16) and (18) to $(\texttt{NIL}, \texttt{NIL})$, and Rule (14) solves the goal.
- If $u = \texttt{NIL}$ and $v = \texttt{CONS}\ N_1\ N_2$, the pair reduces by equations (16), (17) and (19) to

  $$\begin{array}{l}(\texttt{CONS}\,(f\ N_1)\,(\texttt{Lmap}\ f\,(\texttt{Lappend}\,\texttt{NIL}\ N_2)),\\ \ \texttt{CONS}\,(f\ N_1)\,(\texttt{Lappend}\,\texttt{NIL}\,(\texttt{Lmap}\ f\ N_2))).\end{array}$$

  Using equation (16) to replace the second NIL by $\texttt{Lmap}\ f\ \texttt{NIL}$ restores the form of the bisimulation, so that Rule (15) can conclude this case.
- If $u = \texttt{CONS}\ M_1\ M_2$, the pair reduces by equations (17) and (20) to

  $$\begin{array}{l}(\texttt{CONS}\,(f\ M_1)\,(\texttt{Lmap}\ f\,(\texttt{Lappend}\ M_2\ v)),\\ \ \texttt{CONS}\,(f\ M_1)\,(\texttt{Lappend}\,(\texttt{Lmap}\ f\ M_2)\,(\texttt{Lmap}\ f\ v))).\end{array}$$

  Now Rule (15) solves the goal, proving the distributive law.

Two easy theorems state that NIL is the identity element for Lappend. For $M \in \texttt{LList}\ A$ we have

$$\texttt{Lappend}\,\texttt{NIL}\ M = M \quad\text{and}\quad \texttt{Lappend}\ M\ \texttt{NIL} = M \tag{27}$$

Both proofs have only two cases because $M$ is the only variable.

The strong coinduction rule (13) can prove the distributive law with only two cases. Recall from §6.5 that the latter rule implicitly includes the equality relation in the bisimulation; if we can reduce the pair to the form $(a, a)$, then we are done. The simpler proof applies the strong coinduction rule (13), using the same bisimulation as before. Now we must show

$$\begin{aligned}&(\texttt{Lmap}\ f\ (\texttt{Lappend}\ u\ v),\\ &\ \texttt{Lappend}\,(\texttt{Lmap}\ f\ u)\,(\texttt{Lmap}\ f\ v))\ \in\ \texttt{LListD_Fun}(\texttt{diag}\,U)r \cup \texttt{diag}(\texttt{LList}\ U).\end{aligned}$$

There are two cases, considering the form of $u$. If $u = \texttt{CONS}\ M_1\ M_2$ then reason exactly as in the previous proof. If $u = \texttt{NIL}$, the goal reduces by (16) and (27) to

$$(\texttt{Lmap}\ f\ v, \texttt{Lmap}\ f\ v)\ \in\ \texttt{LListD_Fun}(\texttt{diag}\,U)r \cup \texttt{diag}(\texttt{LList}\ U).$$

The pair belongs to $\texttt{diag}(\texttt{LList}\ U)$ because $\texttt{Lmap}\ f\ v \in \texttt{LList}\ U$.

**Typing rules.** Most of our coinduction examples prove equations. But coinduction can also prove typing facts such as $\texttt{Lappend}\ M\ N \in \texttt{LList}\ A$. Again, strong coinduction works better for `Lappend` than weak coinduction. The weak rule (7), requires case analysis on both arguments of `Lappend`; three cases must be considered. The strong rule (8) requires only case analysis on the first argument. The two proofs are closely analogous to those of the distributive law.

**Associativity.** A classic example of structural induction is the associativity of append:

$$\texttt{Lappend}\,(\texttt{Lappend}\ M_1\ M_2)\ M_3 = \texttt{Lappend}\ M_1\ (\texttt{Lappend}\ M_2\ M_3)$$

With weak coinduction, the proof is no longer trivial; it involves a bisimulation in three variables and the proof consists of four cases: `NIL-NIL-NIL`, `NIL-NIL-CONS`, `NIL-CONS`, and `CONS`. Strong coinduction with the bisimulation

$$\{(\texttt{Lappend}\,(\texttt{Lappend}\ u\ M_2)\ M_3,\ \texttt{Lappend}\ u\,(\texttt{Lappend}\ M_2\ M_3))\}_{u\in\texttt{LList}\ A}$$

accomplishes the proof easily. There are only two cases. If $u = \texttt{CONS}\ M\ M'$ then the pair reduces to

$$\begin{aligned}&(\texttt{CONS}\ M\ (\texttt{Lappend}\,(\texttt{Lappend}\ M'\ M_2)\ M_3),\\ &\ \texttt{CONS}\ M\ (\texttt{Lappend}\ M'\ (\texttt{Lappend}\ M_2\ M_3)))\end{aligned}$$

and Rule (15) applies as usual. If $u = \texttt{NIL}$ then both components collapse to $\texttt{Lappend}\ M_2\ M_3$, and the pair belongs to the diagonal set.

## 8.6 A comparison with LCF

Scott's Logic for Computable Functions (LCF) is ideal for reasoning about lazy data structures.[5] Types denote domains and function symbols denote continuous functions. Strict and lazy recursive data types can be defined. Domains contain the bottom element $\perp$, which is the denotation of a divergent computation. Objects can be partial, with components that equal $\perp$. A function $f$ can be partial, with $f\ x = \perp$ for some $x$. A lazy list can be partial, with its head, tail or some elements equal to $\perp$. The relation $x \sqsubseteq y$, meaning "$x$ is less defined or equal to $y$," compares partial objects.

[5] Scott's 1969 paper, which laid the foundations of domain theory, has finally been published [30]. Edinburgh LCF [15], a highly influential system, implemented Scott's logic. Still in print is my account of a successor LCF system [22]. Isabelle provides two versions of LCF: one built upon first-order logic, the other built upon Isabelle/HOL.

LCF's fixedpoint operator expresses recursive objects, including partial functions and recursive lists. The fixedpoint induction rule reasons about the unwinding of recursive objects. It subsumes the familiar structural induction rules and even generalizes them to reason about infinite objects, when the induction formula is chain-complete. Since all formulae of the form $x \sqsubseteq y$ and $x = y$ are chain-complete, LCF can prove equations about lazy lists.

Can our HOL treatment of lazy lists compare with LCF's? The Isabelle theory defines the new type $\alpha$ `llist` to contain the elements of `LList(range(Leaf))`, replacing the clumsy set reasoning by automatic type checking. We can still define lists by corecursion and prove equations by coinduction.

The resulting theory of lazy lists is superficially similar to LCF's. But it lacks general recursion and cannot handle divergent computations. Leclerc and Paulin-Mohring's construction of the prime numbers in Coq [17] reflects these problems. Corecursion cannot express the filter functional, needed for the Sieve of Eratosthenes.

A direct comparison between fixedpoint induction and coinduction is difficult. The rules differ greatly in form and their area of overlap is small. Fixedpoint induction can reason about recursive programs at an abstract level, for instance to prove equivalence of partial functions. Coinduction can prove the equivalence of infinite processes.

I prefer LCF for problems that can be stated entirely in domain theory. But LCF is a restrictive framework: all types must denote domains; all functions must be continuous. HOL is a general logic, free of such restrictions, and yet capable of handling a substantial part of the theory of lazy lists.

# 9 Conclusions

This mechanized theory is a comprehensive treatment of recursive data structures in higher-order logic, generalizing Melham's theory [18]. Melham's approach is concrete: one particular tree structure represents all recursive types. My fixedpoint approach is more abstract, which facilitates extensions such as mutual recursion [26] and infinite branching trees. Users may also add new monotone operators to the type definition language. Types need not be free (such that each constructor function is injective); for example, we may define the set `Fin` $A$ of all finite subsets of $A$ as a least fixedpoint:

$$\texttt{Fin}\ A \equiv \texttt{lfp}\left(\lambda Z.\{\{\}\} \cup \left(\bigcup_{y\in Z}\bigcup_{x\in A}\{\{x\}\cup y\}\right)\right)$$

Most importantly, the theory justifies non-WF data structures.

Elsa Gunter [16] has independently developed a theory of trees, using ideas similar to those of §4. Her aim is to extend Melham's package with infinite branching, rather than coinduction.

What about set theory? The ordered pair $(a, b)$ is traditionally defined to be $\{\{a\}, \{a, b\}\}$. Non-WF data structures presuppose non-WF sets, with infinite descents along the $\in$ relation. Such sets are normally forbidden by the Foundation Axiom, but the Anti-Foundation Axiom (AFA) asserts their existence. Aczel [3] has analysed and advocated this axiom; some authors, such as Milner and Tofte [21], have suggested formalizing coinductive arguments using it.

But non-WF lists and trees are easily expressed in set theory without new axioms. Simply use the ideas presented above for Isabelle/HOL. A new definition of ordered pair, based upon the old one, allows infinite descents. Define the variant ordered pair $(a;b)$ to be $\{(0,x)\}_{x\in a}\cup\{(1,y)\}_{y\in b}$. This is equivalent to the disjoint sum $a+b$ as usually defined, but note also its similarity to $M\cdot N$ (see §4.4). Elsewhere [27] I have developed this approach; it handles recursive data structures in full generality, but not the models of concurrency that motivated Aczel. The proof of the main theorem is inspired by that of Lemma 2. Isabelle/ZF mechanizes this theory.

The theory of coinduction requires further development. Its treatment of lazy lists is clumsy compared with LCF's. Stronger principles of coinduction and corecursion might help. Its connection with similar work in stronger type theories [12, 17] deserves investigation. Leclerc and Paulin-Mohring [17] remark that Coq can express finite and infinite data structures in a manner strongly reminiscent of the Knaster-Tarski Theorem. They proceed to consider a representation of streams that specifies the possible values of the stream member at position $i$, where $i$ is a natural number. Generalizing this approach to other infinite data structures requires generalizing the notion of position, perhaps as in §4.2.

Jacob Frost [11] has performed Milner and Tofte's coinduction example [21] using Isabelle/HOL and Isabelle/ZF. The most difficult task is not proving the theorem but formalizing the paper's non-WF definitions. As of this writing, Isabelle/HOL provides no automatic means of constructing the necessary definitions and proofs.

**Acknowledgements.** Jacob Frost, Sara Kalvala and the referees commented on the paper. Martin Coen helped to set up the environment for coinduction proofs. Andrew Pitts gave much advice, for example on proving that equality is a `gfp`. Tobias Nipkow and his colleagues have made substantial contributions to Isabelle. The research was funded by the SERC (grants GR/G53279, GR/H40570) and by the ESPRIT Basic Research Actions 3245 'Logical Frameworks' and 6453 'Types'.